\pdfoutput=1
\documentclass[12pt,a4paper]{article}

\usepackage{ifthen} 
\newboolean{pdflatex}
\usepackage{tikz}       
\usepackage{lipsum} 
\setboolean{pdflatex}{true} 

\newboolean{articletitles}
\setboolean{articletitles}{true} 

\newboolean{uprightparticles}
\setboolean{uprightparticles}{false} 

\def\paperauthors{LHCb collaboration} 
\def\paperasciititle{Search for GeV-scale axion-like particles in
PbPb collisions} 
\def\papertitle{Search for \gev-scale axion-like particles in
PbPb collisions} 
\def\paperkeywords{{High Energy Physics}, {LHCb}} 
\def\papercopyright{\the\year\ CERN for the benefit of the LHCb collaboration} 
\def\paperlicence{CC BY 4.0 licence}
\def\paperlicenceurl{https://creativecommons.org/licenses/by/4.0/}

\newif\ifEnableSectionTOCLinks
\EnableSectionTOCLinksfalse 

\usepackage[top=1in, bottom=1.25in, left=1in, right=1in]{geometry}

\usepackage{microtype}
\usepackage{lineno}  
\usepackage{xspace} 
\usepackage{caption} 

\usepackage{graphicx}  
\usepackage{color}
\usepackage{colortbl}
\graphicspath{{./figs/}} 

\usepackage{amsmath} 
\usepackage{amssymb}
\usepackage{amsfonts}
\usepackage{upgreek} 

\newcommand*\patchAmsMathEnvironmentForLineno[1]{%
\expandafter\let\csname old#1\expandafter\endcsname\csname #1\endcsname
\expandafter\let\csname oldend#1\expandafter\endcsname\csname
end#1\endcsname
 \renewenvironment{#1}%
   {\linenomath\csname old#1\endcsname}%
   {\csname oldend#1\endcsname\endlinenomath}%
}
\newcommand*\patchBothAmsMathEnvironmentsForLineno[1]{%
  \patchAmsMathEnvironmentForLineno{#1}%
  \patchAmsMathEnvironmentForLineno{#1*}%
}
\AtBeginDocument{%
\patchBothAmsMathEnvironmentsForLineno{equation}%
\patchBothAmsMathEnvironmentsForLineno{align}%
\patchBothAmsMathEnvironmentsForLineno{flalign}%
\patchBothAmsMathEnvironmentsForLineno{alignat}%
\patchBothAmsMathEnvironmentsForLineno{gather}%
\patchBothAmsMathEnvironmentsForLineno{multline}%
\patchBothAmsMathEnvironmentsForLineno{eqnarray}%
}

\usepackage[pdftex,
            pdfauthor={\paperauthors},
            pdftitle={\paperasciititle},
            pdfkeywords={\paperkeywords}]{hyperref}
\usepackage{hyperxmp}
\hypersetup{
    pdfcopyright={Copyright (C) \papercopyright},
    pdflicenseurl={\paperlicenceurl}
}

\usepackage[colorinlistoftodos,textsize=scriptsize]{todonotes}

\usepackage[bottom,flushmargin,hang,multiple]{footmisc}

\usepackage[all]{hypcap} 

\usepackage{xspace} 
\usepackage{upgreek}

\def\lhcb   {\mbox{LHCb}\xspace}
\def\atlas  {\mbox{ATLAS}\xspace}
\def\cms    {\mbox{CMS}\xspace}

\def\belletwo {\mbox{Belle~II}\xspace}
\def\besiii {\mbox{BESIII}\xspace}

\def\lhc    {\mbox{LHC}\xspace}
\def\lep    {\mbox{LEP}\xspace}

\def\velo   {VELO\xspace}

\def\herschel {\mbox{\textsc{HeRSCheL}}\xspace}
\def\spd    {SPD\xspace}

\def\ecal   {ECAL\xspace}

\def\MagUp {\mbox{\em Mag\kern -0.05em Up}\xspace}

\ifdefined\ifuprightparticles
\else
\newboolean{uprightparticles}
\setboolean{uprightparticles}{false} 
\fi

\ifthenelse{\boolean{uprightparticles}}%
{

 \def\Peta        {\ensuremath{\upeta}\xspace}

 \def\Pmu         {\ensuremath{\upmu}\xspace}

 \def\Ppi         {\ensuremath{\uppi}\xspace}

 \def\Ppsi        {\ensuremath{\uppsi}\xspace}

 \def\PDelta      {\ensuremath{\Delta}\xspace}                 
 \def\PXi         {\ensuremath{\Xi}\xspace}                 
 \def\PLambda     {\ensuremath{\Lambda}\xspace}                 
 \def\PSigma      {\ensuremath{\Sigma}\xspace}                 
 \def\POmega      {\ensuremath{\Omega}\xspace}                 
 \def\PUpsilon    {\ensuremath{\Upsilon}\xspace}
 \let\oldPi\Pi
 \def\PPi         {\ensuremath{\oldPi}\xspace}

 \def\PB      {\ensuremath{\mathrm{B}}\xspace}                 
 \def\PD      {\ensuremath{\mathrm{D}}\xspace}                 
 \def\PJ      {\ensuremath{\mathrm{J}}\xspace}                 
 \def\PK      {\ensuremath{\mathrm{K}}\xspace}                 
 \def\Pb      {\ensuremath{\mathrm{b}}\xspace}                 
 \def\Pc      {\ensuremath{\mathrm{c}}\xspace}                 
                  
 \def\Pe      {\ensuremath{\mathrm{e}}\xspace}                 
 \def\Ps      {\ensuremath{\mathrm{s}}\xspace}

 \def\thebaroffset{0.0em}
}
{

 \def\Peta        {\ensuremath{\eta}\xspace}

 \def\Pmu         {\ensuremath{\mu}\xspace}

 \def\Ppi         {\ensuremath{\pi}\xspace}

 \def\Ppsi        {\ensuremath{\psi}\xspace}                 
                  
 \mathchardef\PDelta="7101
 \mathchardef\PXi="7104
 \mathchardef\PLambda="7103
 \mathchardef\PSigma="7106
 \mathchardef\POmega="710A
 \mathchardef\PUpsilon="7107
 \mathchardef\PPi="7105
 \def\PB      {\ensuremath{B}\xspace}                 
 \def\PD      {\ensuremath{D}\xspace}                 
 \def\PJ      {\ensuremath{J}\xspace}                 
 \def\PK      {\ensuremath{K}\xspace}                 
 \def\Pb      {\ensuremath{b}\xspace}                 
 \def\Pc      {\ensuremath{c}\xspace}                 
                  
 \def\Pe      {\ensuremath{e}\xspace}                 
 \def\Ps      {\ensuremath{s}\xspace}

 \def\thebaroffset{0.18em}
}
\newcommand{\offsetoverline}[2][\thebaroffset]{\kern #1\overline{\kern -#1 #2}}%

\makeatletter
\ifcase \@ptsize \relax
  \newcommand{\miniscule}{\@setfontsize\miniscule{4}{5}}
\or
  \newcommand{\miniscule}{\@setfontsize\miniscule{5}{6}}
\or
  \newcommand{\miniscule}{\@setfontsize\miniscule{5}{6}}
\fi
\makeatother

\DeclareRobustCommand{\optbar}[1]{\shortstack{{\miniscule (\rule[.5ex]{1.25em}{.18mm})}
  \\ [-.7ex] $#1$}}

\def\en         {{\ensuremath{\Pe^-}}\xspace}   
\def\ep         {{\ensuremath{\Pe^+}}\xspace}

\def\mup        {{\ensuremath{\Pmu^+}}\xspace}
\def\mun        {{\ensuremath{\Pmu^-}}\xspace} 

\def\squark    {{\ensuremath{\Ps}}\xspace}

\def\cquark    {{\ensuremath{\Pc}}\xspace}

\def\bquark    {{\ensuremath{\Pb}}\xspace}

\def\pion   {{\ensuremath{\Ppi}}\xspace}
\def\piz    {{\ensuremath{\pion^0}}\xspace}

\def\kaon    {{\ensuremath{\PK}}\xspace}

\def\KorKbar {\kern \thebaroffset\optbar{\kern -\thebaroffset \PK}{}\xspace}

\def\Kp      {{\ensuremath{\kaon^+}}\xspace}

\def\Kstarp  {{\ensuremath{\kaon^{*+}}}\xspace}

\def\D       {{\ensuremath{\PD}}\xspace}

\def\DorDbar {\kern \thebaroffset\optbar{\kern -\thebaroffset \PD}\xspace}

\def\Dp      {{\ensuremath{\D^+}}\xspace}
\def\Dm      {{\ensuremath{\D^-}}\xspace}

\def\DpDm    {\ensuremath{\Dp {\kern -0.16em \Dm}}\xspace}

\def\B       {{\ensuremath{\PB}}\xspace}

\def\BorBbar {\kern \thebaroffset\optbar{\kern -\thebaroffset \PB}\xspace}

\def\Bd      {{\ensuremath{\B^0}}\xspace}

\def\BdorBdbar {\kern \thebaroffset\optbar{\kern -\thebaroffset \Bd}\xspace}
\def\Bu      {{\ensuremath{\B^+}}\xspace}

\def\Bp      {{\ensuremath{\Bu}}\xspace}

\def\Bs      {{\ensuremath{\B^0_\squark}}\xspace}

\def\BsorBsbar {\kern \thebaroffset\optbar{\kern -\thebaroffset \Bs}\xspace}

\def\jpsi     {{\ensuremath{{\PJ\mskip -3mu/\mskip -2mu\Ppsi}}}\xspace}

\def\Y#1S{\ensuremath{\PUpsilon{(#1S)}}\xspace}

\def\LorLbar     {\kern \thebaroffset\optbar{\kern -\thebaroffset \PLambda}\xspace}

\newcommand{\decay}[2]{\ensuremath{\mathinner{#1\!\to #2}}\xspace}

\def\to                 {\ensuremath{\rightarrow}\xspace}

\def\AT#1     {\ensuremath{A_{\mathrm{T}}^{#1}}\xspace}           

\def\C#1      {\ensuremath{\mathcal{C}_{#1}}\xspace}                       
\def\Cp#1     {\ensuremath{\mathcal{C}_{#1}^{'}}\xspace}                    
\def\Ceff#1   {\ensuremath{\mathcal{C}_{#1}^{\mathrm{(eff)}}}\xspace}        
\def\Cpeff#1  {\ensuremath{\mathcal{C}_{#1}^{'\mathrm{(eff)}}}\xspace}       
\def\Ope#1    {\ensuremath{\mathcal{O}_{#1}}\xspace}                       
\def\Opep#1   {\ensuremath{\mathcal{O}_{#1}^{'}}\xspace}                    

\newcommand{\aunit}[1]{\ensuremath{\text{\,#1}}}       

\newcommand{\tev}{\aunit{Te\kern -0.1em V}\xspace}
\newcommand{\gev}{\aunit{Ge\kern -0.1em V}\xspace}
\newcommand{\mev}{\aunit{Me\kern -0.1em V}\xspace}
\newcommand{\kev}{\aunit{ke\kern -0.1em V}\xspace}
\newcommand{\ev}{\aunit{e\kern -0.1em V}\xspace}
\newcommand{\gevgev}{\ensuremath{\gev^2}\xspace} 
\newcommand{\mevc}{\ensuremath{\aunit{Me\kern -0.1em V\!/}c}\xspace}
\newcommand{\gevc}{\ensuremath{\aunit{Ge\kern -0.1em V\!/}c}\xspace}
\newcommand{\mevcc}{\ensuremath{\aunit{Me\kern -0.1em V\!/}c^2}\xspace}
\newcommand{\gevcc}{\ensuremath{\aunit{Ge\kern -0.1em V\!/}c^2}\xspace}

\def\m    {\aunit{m}\xspace}

\def\nb {\aunit{nb}\xspace}
\def\invnb {\ensuremath{\nb^{-1}}\xspace}
\def\pb {\aunit{pb}\xspace}

\def\gsim{{~\raise.15em\hbox{$>$}\kern-.85em
          \lower.35em\hbox{$\sim$}~}\xspace}
\def\lsim{{~\raise.15em\hbox{$<$}\kern-.85em
          \lower.35em\hbox{$\sim$}~}\xspace}

\def\PDF {PDF\xspace}

\def\sqsnn {\ensuremath{\protect\sqrt{s_{\scriptscriptstyle\text{NN}}}}\xspace}

\def\ptsq       {\ensuremath{p_{\mathrm{T}}^2}\xspace}

\def\et         {\ensuremath{E_{\mathrm{T}}}\xspace}

\def\evtgen     {\mbox{\textsc{EvtGen}}\xspace}

\def\geant      {\mbox{\textsc{Geant4}}\xspace}

\def\photos     {\mbox{\textsc{Photos}}\xspace}

\def\starlight  {\mbox{\textsc{STARlight}}\xspace}
\def\superchic  {\mbox{\textsc{SuperChic}}\xspace}

\def\tell1  {TELL1\xspace}
\def\ukl1   {UKL1\xspace}

\newcommand{\eg}{\mbox{\itshape e.g.}\xspace}

\newcommand{\lhcborcid}[1]{\href{https://orcid.org/#1}{\hspace*{0.1em}\raisebox{-0.45ex}{\includegraphics[width=1em]{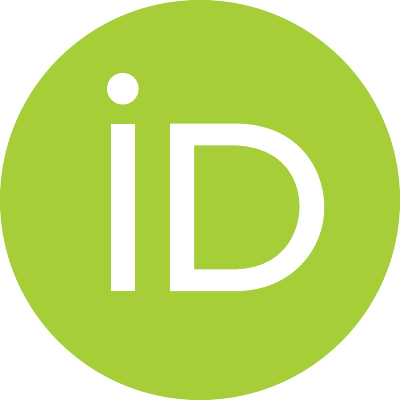}}}}

\hypersetup{
  colorlinks   = true, 
  urlcolor     = blue, 
  linkcolor    = blue, 
  citecolor    = red   
}

\ifEnableSectionTOCLinks
    \usepackage[explicit]{titlesec} 
    
    \let\oldcontentsline\contentsline
    \renewcommand

    \titleformat{\section}{\normalfont\Large\bf}{\hyperlink{tocsection.\thesection}{{\thesection} \parbox[t]{\dimexpr\textwidth-1pc}{#1}}}{1pc}{}

    \titleformat{\subsection}{\normalfont\bf}{\hyperlink{tocsubsection.\thesubsection}{{\thesubsection} \parbox[t]{\dimexpr\textwidth-1pc}{#1}}}{1pc}{}

    \titleformat{name=\section,numberless}[display]{}{}{0pt}{\normalfont\Huge\bfseries #1}
\fi

\usepackage{cite} 
\usepackage{LHCb/mciteplus}

\usepackage{longtable} 

\def\ggF {ggF\xspace}
\def\LbL {LbL\xspace}
\def\BWh {BWh\xspace}
\def\ALP {ALP\xspace}
\def\ALPs {ALPs\xspace}

\def\JPsiee  {\decay{\jpsi}{\ep\en}}
\def\galp    {\ensuremath{g_{a\gamma}}\xspace}
\def\malp    {\ensuremath{m_{a}}\xspace}

\begin{document}

\renewcommand{\thefootnote}{\fnsymbol{footnote}}
\setcounter{footnote}{1}


\begin{titlepage}
\pagenumbering{roman}

\vspace*{-1.5cm}
\centerline{\large EUROPEAN ORGANIZATION FOR NUCLEAR RESEARCH (CERN)}
\vspace*{1.5cm}
\noindent
\begin{tabular*}{\linewidth}{lc@{\extracolsep{\fill}}r@{\extracolsep{0pt}}}
\ifthenelse{\boolean{pdflatex}}
{\vspace*{-1.5cm}\mbox{\!\!\!\includegraphics[width=.14\textwidth]{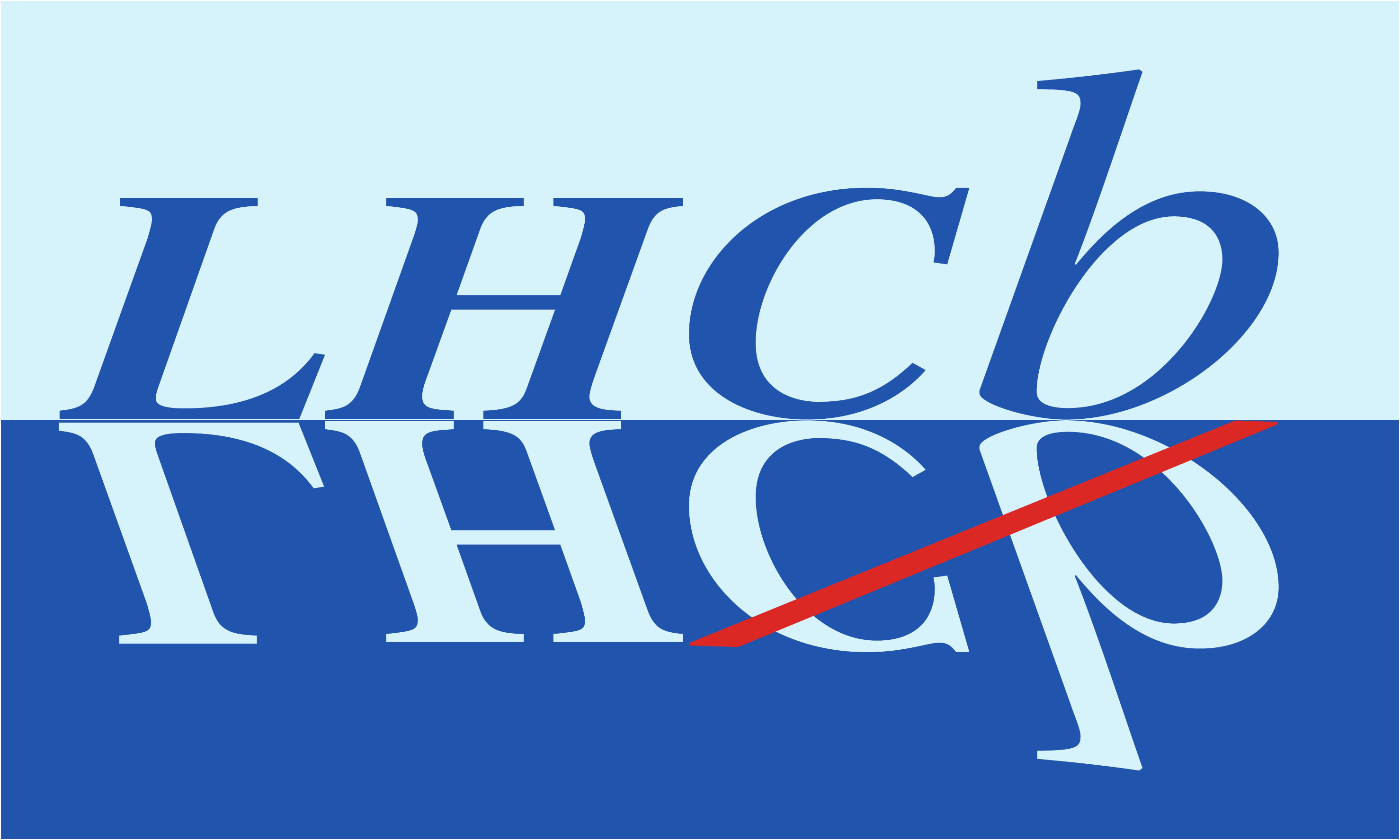}} & &}%
{\vspace*{-1.2cm}\mbox{\!\!\!\includegraphics[width=.12\textwidth]{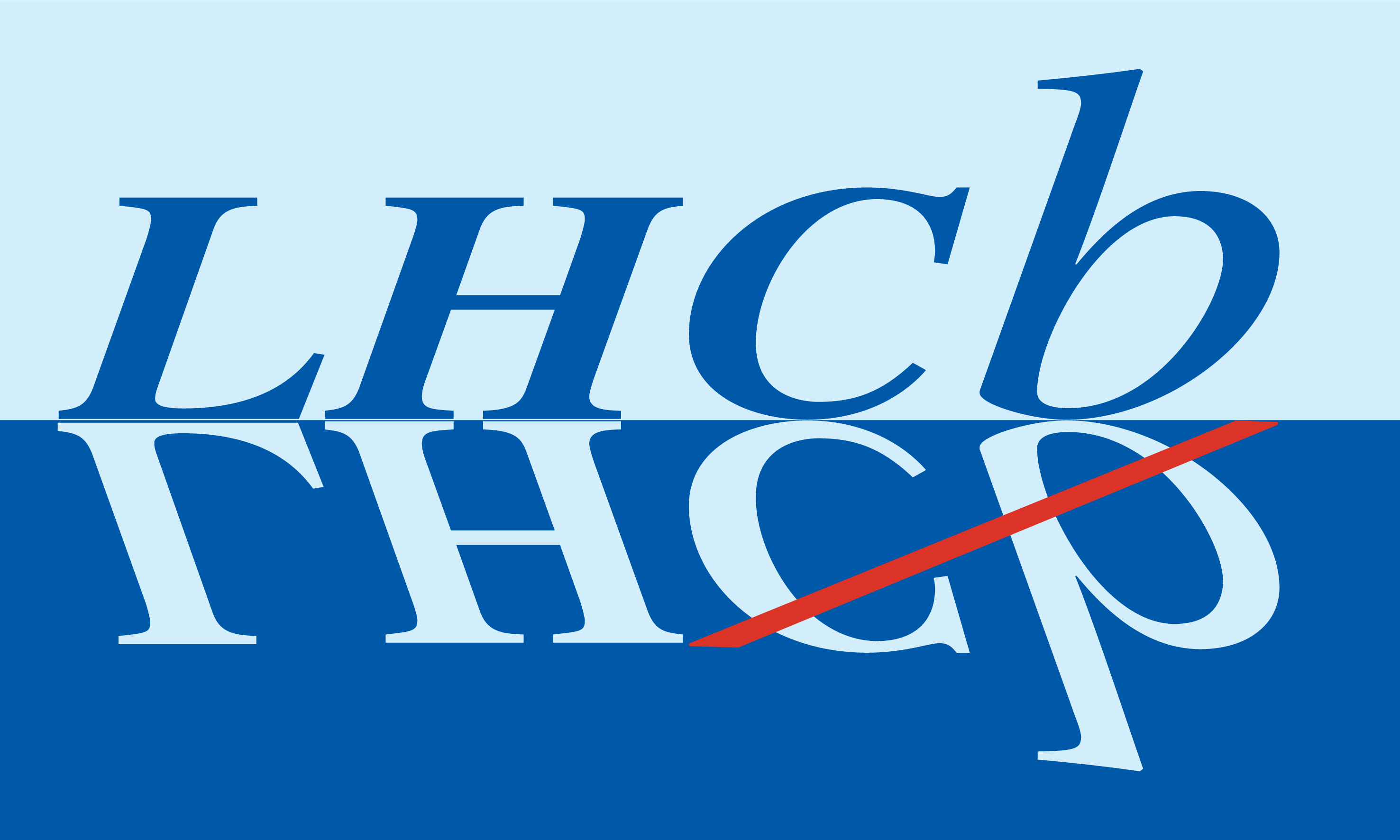}} & &}%
\\
 & & CERN-EP-2026-273 \\  
 & & LHCb-PAPER-2026-029\\  
 & & October 29, 2026 \\ 
 & & \\
\end{tabular*}

\vspace*{4.0cm}

{\normalfont\bfseries\boldmath\huge
\begin{center}
  \papertitle 
\end{center}
}

\vspace*{2.0cm}

\begin{center}
\paperauthors\footnote{Authors are listed at the end of this paper.}
\end{center}

\vspace{\fill}

\begin{abstract}
  \noindent
A search is performed for the production of axion-like particles with masses in the range 3--10\gev using  PbPb collision data recorded by the LHCb experiment, corresponding to an integrated luminosity of 0.228\invnb and a nucleon-nucleon centre-of-mass energy of $\sqsnn=5.02\tev$.
This is the first search for axion-like particles produced exclusively by two photons ($\gamma\gamma \rightarrow a \rightarrow \gamma\gamma$) in the forward direction. 
It extends sensitivity to previously unexplored regions of the parameter space, probing axion-like particle masses and couplings down to 3.0\gev and $3.2\times10^{-4}\gev^{-1}$, respectively. 
This result provides the most stringent limits to date on the production of axion-like particles from exclusive photon-photon fusion processes with masses between 3 and 5\gev. 
  
\end{abstract}

\vspace*{2.0cm}

\begin{center}
  Submitted to JHEP
\end{center}

\vspace{\fill}

{\footnotesize 
\centerline{\copyright~\papercopyright. \href{\paperlicenceurl}{\paperlicence}.}}
\vspace*{2mm}

\end{titlepage}


\newpage
\setcounter{page}{2}
\mbox{~}
%
%
%
%


\renewcommand{\thefootnote}{\arabic{footnote}}
\setcounter{footnote}{0}


\cleardoublepage


\pagestyle{plain} 
\setcounter{page}{1}
\pagenumbering{arabic}


\section{Introduction}
\label{sec:Introduction}
Many extensions of the Standard Model predict new (pseudo)scalar particles, including those that share the quantum numbers of axions~\cite{PhysRevLett.38.1440,PhysRevLett.40.223}, generically referred to as axion-like particles (\ALPs), $a$. 
These particles are pseudo Nambu-Goldstone bosons arising from a spontaneously broken approximate symmetry~\cite{Ringwald:2014vqa,PhysRevLett.38.1440,PhysRevLett.40.279}, analogous to neutral pions arising  from the approximate chiral symmetry of quantum chromodynamics (QCD)~\cite{Koch:1997ei}. 
Axion-like particles appear naturally in many beyond-the-Standard-Model (BSM) scenarios, for example in portals to the dark sector~\cite{Nomura:2009,Dolan_2015,Kozaczuk_2015,Kaneta:2017wfh,Kamada:2017tsq,Gola:2021abm,PhysRevD.108.075003,Armando:2023zwz}, or models of the QCD axion~\cite{PhysRevLett.38.1440,PhysRevLett.40.223,PhysRevLett.40.279}  or dark-matter candidates~\cite{Duffy_2009,ABBOTT1983133,PRESKILL1983127,DINE1983137}. 
They also arise in Higgs compositeness models~\cite{Cacciapaglia_2019}, extended Higgs sectors~\cite{Branco_2012}, string theory frameworks~\cite{Ringwald:2012cu,Svrcek:2006yi,Witten:1984dg,Arvanitaki:2009fg} and in the form of $R$-axions from dynamical supersymmetry breaking~\cite{Bagger:1994hh,Frere:1983ag,Nelson:1993nf,Bellazzini:2017neg}.

Searches for \ALPs have been carried out in recent years using ultraperipheral collisions (UPC) in lead-lead (PbPb) data at the \lhc  \cite{ATLAS:2020hii,CMS:2018erd,CMS:2024bnt}. 
In these, the impact parameter is larger than the sum of the nuclear radii, resulting in an extremely high photon flux~\cite{Baltz:2007kq}---proportional to the square of the ion's atomic number---that enhances the two-photon interaction rate  substantially~\cite{dEnterria:2013zqi}. 
This environment enables searches for (pseudo)scalars that only couple to photons, enabling the process $\gamma\gamma \to a \to \gamma\gamma$ to be investigated~\cite{dEnterria:2021ljz}. 

Previous limits on \ALPs coupling to photons have been set by experiments \eg at colliders \cite{OPAL:2002vhf,Knapen:2016moh,Jaeckel:2015jla,Babar,BESIII},  
accelerators \cite{PrimEx:2010fvg,Aloni:2019ruo, GlueX}, 
beam dumps \cite{beamdump, beamdump2} 
and using data from low-energy supernovae~\cite{Low-E}.
For \ALP masses $\malp < 1\gev$,\footnote{Natural units are used throughout this paper.} upper limits are still dominated by these earlier measurements, while above 5\gev the most stringent limits are set by the \atlas~\cite{ATLAS:2020hii} and \cms~\cite{CMS:2018erd,CMS:2024bnt} collaborations. 
In the interval $1<\malp<5\gev$, the strongest constraints are provided by \belletwo~\cite{Belle-II:2020jti,Belle-II:2026znr} and \lep data~\cite{OPAL:2002vhf,Knapen:2016moh,Jaeckel:2015jla}; there are no dedicated searches from \lhc experiments. However, phenomenological studies suggest that the PbPb UPC data collected by \lhcb are sensitive  to \ALP masses that are not excluded by  previous measurements~\cite{Coelho:2020saz,Goncalves:2021pdc,Knapen:2016moh,Baldenegro:2019whq}. 

The ability of the \lhcb experiment to trigger on and reconstruct photons with low transverse energy provides unique sensitivity that  complements the reach of \belletwo and \lep constraints in the intermediate-mass region.
This search targets \ALP masses in the range 3--10\gev. The lower boundary is determined by the trigger requirements applied during data taking. In this region, the analysis has an  expected sensitivity that can improve limits from exclusive photon-photon fusion processes relative to those from previous experiments, particularly in the 3--5\gev range.

In the search for resonances decaying into two photons in UPC events, three primary sources of background must be considered: light-by-light scattering, $\gamma\gamma \rightarrow \gamma\gamma$ (LbL); gluon-gluon fusion, $gg \rightarrow \gamma\gamma$ (ggF); and the Breit-Wheeler process, $\gamma\gamma \rightarrow \ep\en$ (BWh)~\cite{Breit:1934zz}. 
In the  3--10\gev mass interval, \JPsiee production in UPC is also accounted for as a source of contamination. 
Both the BWh and \JPsiee channels contribute to the background through electron to photon misidentification.

This paper presents a search for axion-like particle production in the mass range 3--10\gev, using PbPb collision data recorded by the LHCb experiment in 2018 at a nucleon-nucleon centre-of-mass energy of $\sqsnn = 5.02\tev$, corresponding to an integrated luminosity of $0.228\pm0.010\invnb$. During part of the data taking, gas was injected into the interaction region, so that beam-gas collisions can also occur.
In the  scenario considered, the \ALP couples to photons and decays via \decay{a}{\gamma\gamma} with a coupling strength \galp. 
This search extends the mass region previously explored by \lhc experiments and is the first study of this process in the unique forward acceptance of \lhcb.

\section{Experiment and simulation}
\label{sec:experiment}
The \lhcb detector~\cite{LHCb-DP-2008-001,LHCb-DP-2014-002} is a single-arm forward spectrometer covering the pseudorapidity region $2<\eta<5$, designed for the study of particles containing \bquark or \cquark quarks.
The detector includes a high-precision tracking system consisting of a silicon-strip vertex detector surrounding the interaction region (VELO), a large-area silicon-strip detector located upstream of a dipole magnet with a bending power of about $4{\mathrm{\,Tm}}$, and three tracking stations of silicon-strip detectors and straw drift tubes placed downstream of the magnet.
Photons, electrons, and hadrons are identified by a calorimeter system comprising scintillating-pad (\spd) and preshower (PS) detectors, an electromagnetic calorimeter (\ecal), and a hadronic calorimeter.
Muons are identified by a system composed of alternating layers of iron and multiwire proportional chambers.

The \ecal is particularly important for this analysis, consists of alternating layers of lead and scintillator, and provides an energy ($E$) resolution of
$13.5\%/\sqrt{E/\gev{}} \oplus 5.2\% \oplus (0.32\gev{})/E$~\cite{LHCb-DP-2020-001}.
The pseudorapidity coverage is extended by the forward shower counters, the \herschel detector~\cite{LHCb-DP-2016-003}, consisting of five planes of scintillators positioned around the beam pipe. 
Three of these planes are placed upstream of the \lhcb detector at 114\m, 19.7\m and 7.5\m from the nominal interaction point, while the other two are located downstream at 20\m and 114\m.
By covering the pseudorapidity range of $5 \lesssim |\eta| \lesssim 10$, the \herschel detector significantly enhances the ability to identify UPC events by detecting particles from dissociated nucleons that would otherwise escape the \lhcb acceptance.

Simulated data samples are used to model the detector response to diphoton candidates from both signal and background processes.
The \ALP signal samples are generated using the \starlight 3.13~\cite{Klein:2016yzr} event generator for \malp values of 3, 4, 5, 6, 7, 8, 9 and 10\gev.
The background processes BWh and \JPsiee are also generated using \starlight, while the ggF processes are simulated using \superchic 2.04~\cite{superchic4} in the proton-proton ($pp$) configuration.
Decays of unstable particles are handled by \evtgen~\cite{Lange:2001uf}, with final-state radiation simulated using \photos~\cite{davidson2015photos}.
The interaction of all generated particles with the detector material, along with the detector's response, is implemented using the \geant toolkit~\cite{Allison:2006ve,Agostinelli:2002hh}, as described in Ref.~\cite{LHCb-PROC-2011-006}.

As the version of \superchic\ implemented within the LHCb framework does not include LbL production, this process is simulated using  \superchic 5.4.1, with detector response modelled by smearing photon-related kinematic variables using fully simulated ggF events.

\section{Event selection}
\label{sec:selection}

Candidates are selected by online and offline selections. The data used in the search for \ALP candidates were collected using a low-multiplicity hardware trigger that required no more than 50 \spd deposits and at least one photon candidate with  transverse energy $\et > 960\mev$.
The photon candidate is defined as the highest \et cluster in the \ecal with either one or two PS cells hit in front of the cluster and no hits in the corresponding \spd cells.
In the inner region of the \ecal, clusters with three or four PS cells hit in front of them are also accepted as photon candidates.
For the offline selection, the diphoton decay candidate consists of two reconstructed photons each with  $\et > 1300\mev$ in a low track-multiplicity event, having no reconstructed charged particle tracks in the \velo and fewer than ten clusters in the \velo. 
To suppress background from inelastic PbPb collisions, particularly those in which one or both lead nuclei undergo electromagnetic dissociation, a further multiplicity veto requirement that combines the signals in all \herschel stations is imposed~\cite{LHCb-DP-2016-003}.

Photons from \ALP decays are expected to be produced nearly back-to-back in the transverse plane due to momentum conservation, leading to small values of the acoplanarity, $A_{\phi} \equiv 1 - |\Delta_{\phi}|/\pi$, and low diphoton transverse momentum squared, $\ptsq(\gamma\gamma)$.
Here, $\Delta_{\phi}$ is the difference in azimuthal angle between the two photons.
These requirements are essential to suppress background processes having a UPC-like topology, where two photons are produced together with additional particles or photons that escape detection.
The efficiency of these selections depends on the \ALP mass hypothesis.  
To maintain good signal to background ratio, particularly for masses in the range 3--5\gev, the requirements  
\(A_{\phi} < 0.02\) and \(\ptsq(\gamma\gamma) < 0.03\gevgev\) are imposed.

Charged particles that are not reconstructed as tracks but have sufficient momentum to be deflected into the calorimeter acceptance after passing through the magnet, and therefore may be reconstructed as photon candidates, are a potential source of contamination. 
This background is concentrated around the central horizontal region of the calorimeter, as the vertical magnetic field primarily deflects particles in the horizontal plane.
To suppress this, photon candidates are rejected in the
azimuthal regions most affected by requiring $|\phi_{\gamma}/\pi| > 0.1$ and $|\phi_{\gamma}/\pi| < 0.9$.

\section{Determination of signal and background yields}
\label{sec:expected}

The expected signal and background yields are estimated using simulated samples.
The detector acceptance, the photon reconstruction and selection efficiencies, and the diphoton mass-selection efficiency are obtained from simulation, while the remaining efficiencies are evaluated using  data, as described below.
Since the LbL simulation only includes the detector response for the photon kinematic variables, the detector acceptance and photon-reconstruction efficiency for this process are taken from the ggF sample, which is found to be $(10.9 \pm 0.1)\%$.

Both the signal and background processes are characterised by intact lead ions in the final state. 
The suppression of PbPb inelastic collisions, implemented through the \herschel-veto criteria, 
is therefore assigned as a common efficiency for signal and background. 
The working point is chosen to retain an efficiency of approximately 90\%, which is estimated using a pure sample of dimuon BWh events.

The efficiency of the photon-related requirements of the analysis  is determined using an independent online selection.
At the hardware stage, this selection requires at least one and no more than 50 \spd hits.
Events passing this requirement are subsequently processed by the software trigger, which requires at least one \velo track.
To determine the efficiency of the multiplicity requirement imposed by the hardware trigger, events collected randomly during LHC operations having a single beam present in the direction of the LHCb detector are used.
These `beam-empty' events are also used to evaluate the efficiency of other multiplicity requirements.

The photon-related trigger efficiency is determined using a tag-and-probe method similar to that developed in Ref.~\cite{LHCb-DP-2013-002} based on two-photon decays of \Peta and \piz mesons, which are selected using an online trigger requirement that is independent of the photon-related trigger under study.
To avoid overlaps between the photons that could bias the efficiency measurement, a minimum spatial separation between the two photons is imposed~\cite{LHCb-DP-2020-001}. 
The probe photon is chosen randomly and the trigger efficiency is obtained by comparing the number of meson candidates extracted from a mass fit with the number that also satisfies the photon trigger.

\begin{figure}[t]
\begin{center}
\includegraphics[width=0.49\linewidth]{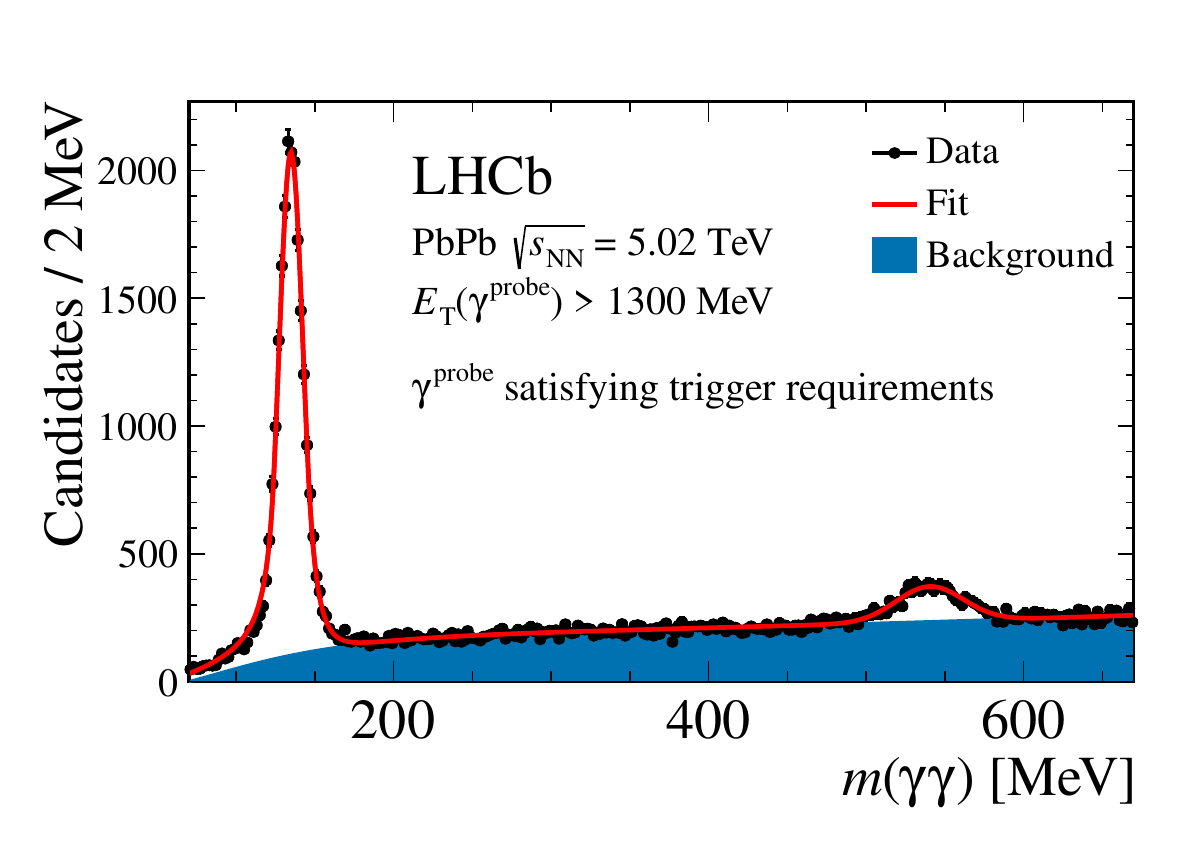}
\includegraphics[width=0.49\linewidth]{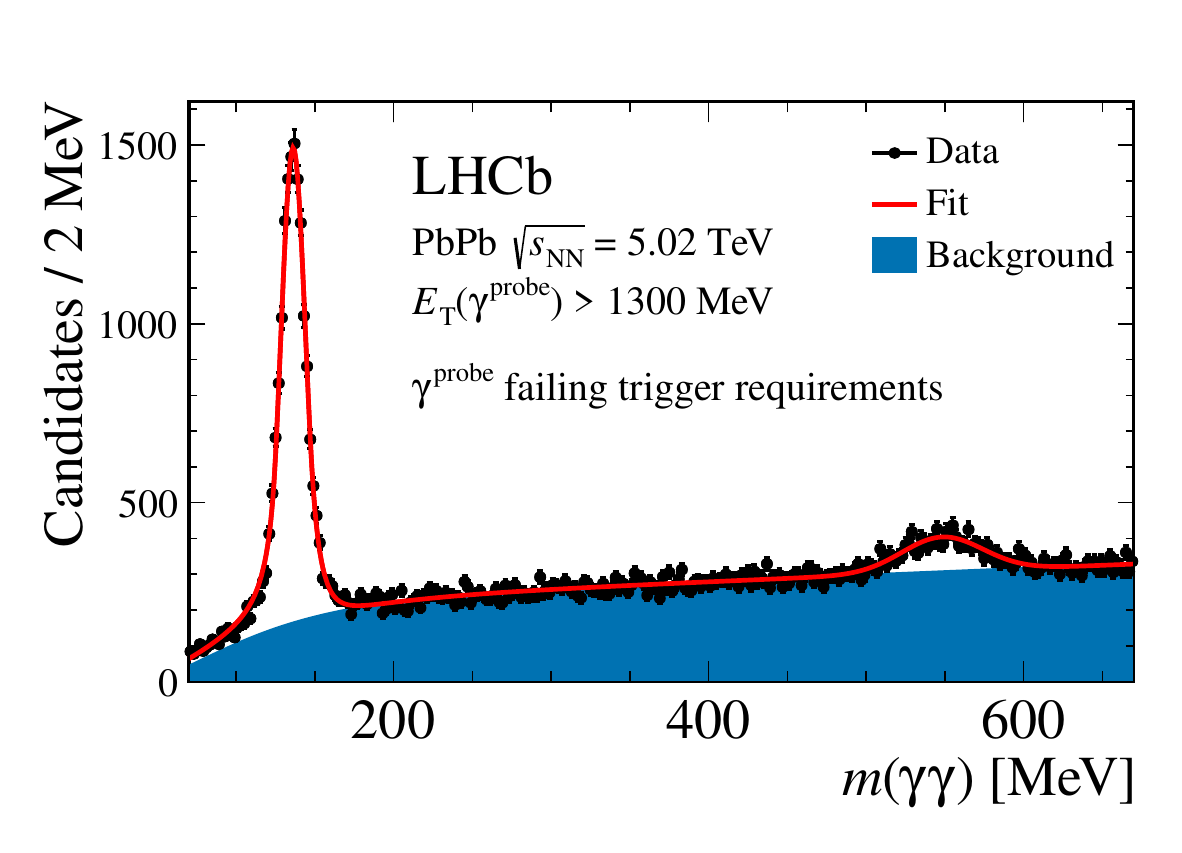}
\vspace*{-0.5cm}
\end{center}
\caption{
Diphoton invariant-mass distribution for the data sample used to determine the trigger efficiencies. The (solid line) total fit includes the background model along with the \piz and \Peta components. The left (right) plot shows the distribution where the probe photon satisfies (fails) the trigger requirements.
}
\label{fig:pizeta}
\end{figure}

The mass distributions of \piz and \Peta candidates are modelled using a Crystal Ball~\cite{Skwarnicki:1986xj} function, while the combinatorial background is described by a smooth threshold function as in Ref.~\cite{LHCb-PAPER-2018-013}.
The trigger efficiency shows no significant dependence on the photon pseudorapidity.
The fit is therefore performed using two samples of diphoton candidates: one where the probe photon passes, and the other where it fails the trigger requirements, as shown in Fig.~\ref{fig:pizeta}.
Above the required $\et(\gamma)$ threshold, the trigger efficiency is measured to be $(58.4 \pm 0.4)\%$.

For the signal, as well as for the LbL and ggF backgrounds, the efficiency associated with the multiplicity requirements on the number of \velo tracks, \velo clusters, \spd hits and reconstructed photons is determined using beam-empty events.
These events provide a good proxy for exclusive two-photon production in the detector and are therefore suitable for evaluating the corresponding efficiencies.
The efficiency for these requirements is defined as the ratio of events passing the selection to the total number of events, and is found to be $(83.3\pm0.3)\%$.

To determine the efficiency associated with the multiplicity requirements on the number of \velo tracks, \velo clusters and additional reconstructed photons for the BWh and \JPsiee background processes, simulated samples of these processes are used. 
No fully simulated events remain in a sample that is approximately ten times larger than data.
Therefore the acoplanarity and \ptsq requirements are removed, while the $\et(\gamma)$ threshold is reduced to 200\mev to enable the study to be performed. 
The resulting efficiency is found to be $(0.02\pm0.01)\%$. 
The efficiency associated with the requirement on the number of \spd hits is taken from beam-empty events and used as an upper limit on this efficiency, which is found to be 99.7\%.

The diphoton invariant-mass selection efficiency for the BWh and \JPsiee background processes is taken from simulation where the requirements on the number of \velo tracks and  clusters are removed, due to finite simulated sample size. 
This approach is considered acceptable as the diphoton invariant-mass distribution shows no dependence on the number of \velo tracks or clusters.

The cross-sections for the \ALP signal and for the LbL, ggF and BWh background processes are calculated with \superchic using the MMHT2015qed Leading Order \PDF set~\cite{Cridge:2023ryv}.
Since the ggF cross-section is computed in the proton-proton configuration, it is scaled by a factor $A^4 R_g^4$, with $A=208$ and $R_g=0.15$~\cite{Guzey:2016tek}, where $A$ is the nucleus mass number and $R_g$ accounts for nuclear shadowing.
Additionally, the probability for producing the diphoton system without accompanying hadronic activity is assumed to scale from $pp$ to PbPb collisions as $\langle S^2 \rangle_{\mathrm{PbPb}} = \langle S^2 \rangle_{pp} / A^2$~\cite{Kaidalov:2003vg,Goncalves:2010dw,Goncalves:2020bqi}.

\begin{figure}[tbh]
\begin{center}
\includegraphics[width=0.70\linewidth]{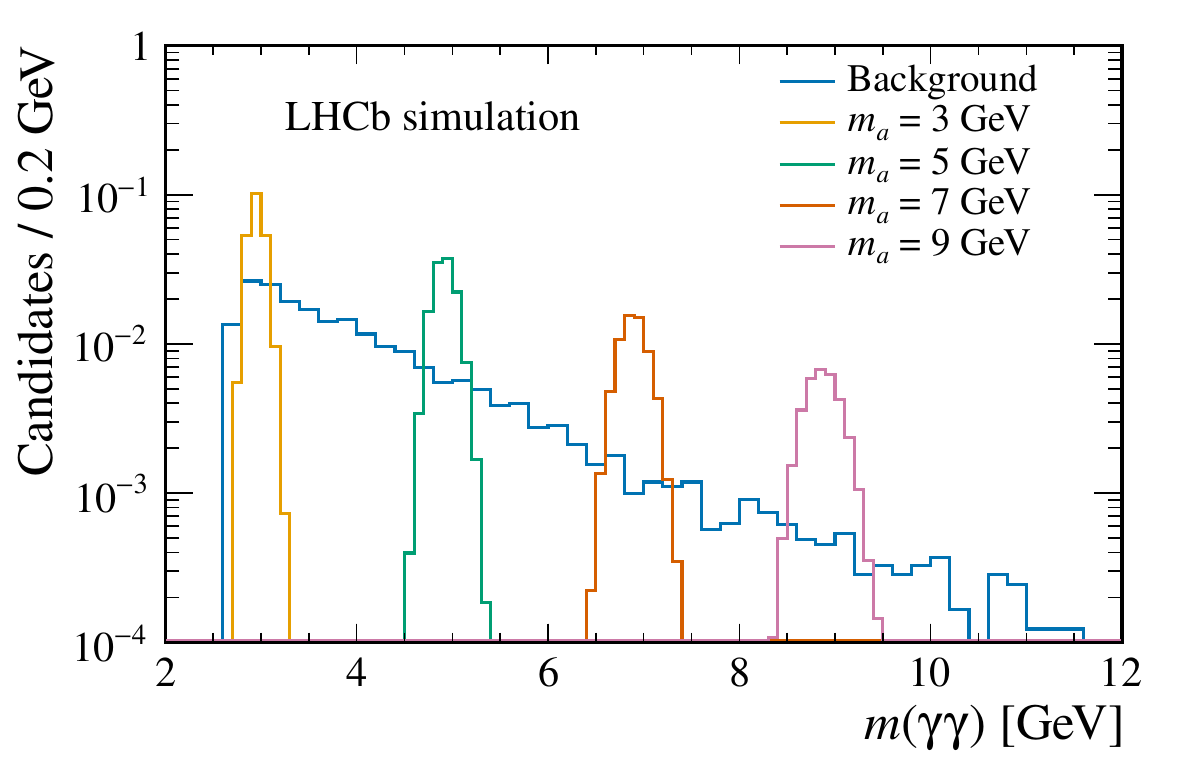}
\vspace*{-0.5cm}
\end{center}
\caption{
Background and signal diphoton invariant-mass distributions, normalised to the expected number of candidates. 
The signal distributions are shown for the 3, 5, 7 and 9\gev  \ALP mass hypotheses, assuming a coupling of $\galp = 10^{-4}\gev^{-1}$.
}
\label{fig:alp_bkg}
\end{figure}

Without applying the diphoton invariant-mass selection, the predicted number of events is 0.2 for the LbL background, $4\times 10^{-4}$ for the ggF background, $3\times 10^{-3}$ for the BWh background, and $2\times 10^{-5}$ for the  \JPsiee background.
No diphoton candidates survive the selection criteria in data.
The expected number of \ALP signal candidates varies from 22.5 to 2.4 for mass hypotheses between 3\gev and 10\gev, assuming a coupling value consistent with existing limits of $\galp = 10^{-3}\gev^{-1}$.
The simulated diphoton invariant-mass distribution is shown in Fig.~\ref{fig:alp_bkg}, where a coupling of $\galp = 10^{-4}\gev^{-1}$ is used to normalise the signal distribution.

To set upper limits on the \ALP mass hypotheses while avoiding potential mismodelling of the diphoton invariant-mass spectrum that could bias a limit determination, the upper limits are evaluated in mass windows centred on the mean reconstructed diphoton mass for each hypothesis. For each \ALP mass point, the diphoton invariant-mass selection window is defined such that it retains 99.7\% of the corresponding signal distribution.

\section{Systematic uncertainties}
\label{sec:systematic}
Systematic uncertainties on the upper limit calculation arise from the efficiency determination, cross-section determination for the background processes, and knowledge of the integrated luminosity.  A summary of the systematic uncertainties is presented in Table~\ref{tab:sys}.

The relative uncertainty on the integrated luminosity is 4.2\% \cite{LHCb-PAPER-2014-047}.
The uncertainty associated with the \herschel efficiency is 1.4\% \cite{LHCb-PAPER-2022-012}.
A common photon reconstruction systematic uncertainty of 4.5\%, determined from \decay{\Bp}{\jpsi\Kstarp} and \decay{\Bp}{\jpsi\Kp} decays~\cite{LHCb-PAPER-2024-015}, is assigned to all samples.

For the \ALP signal, as well as for the LbL and ggF backgrounds, the determination of the efficiency associated with the multiplicity requirements on the number of \velo tracks and clusters, additional reconstructed photons, and \spd hits may be biased due to higher detector activity than expected for events containing only two photons. 
This arises because the efficiencies are evaluated using beam-empty  events, whereas not all data were collected with gas injected into the interaction region~\cite{LHCb-PAPER-2014-047}.
In the presence of gas injection, beam-gas interactions can occur, producing additional particles that may bias multiplicity-related variables and therefore affect the efficiency.
To account for this potential bias, the efficiency is also evaluated using a subset of events that satisfy the \herschel veto requirement, which suppresses events in which one or both lead ions undergo nuclear excitation.
In this case, the efficiency is found to differ by 0.3\% relative to the nominal value; this difference is assigned as a systematic uncertainty.

The uncertainty on the efficiency associated with the multiplicity requirements on the number of \velo tracks, \velo clusters, and additional reconstructed photons is dominated by the limited size of the simulated samples, reaching up to 39\% (50\%) for the BWh (\JPsiee) channels.
The efficiency of the \spd hit requirement is taken from beam-empty event backgrounds. 
This approach may be biased because in BWh and \JPsiee events, two tracks could potentially hit the \spd subdetector. 
To investigate this potential bias, the efficiency is calculated by replacing the requirement of zero \velo tracks with a requirement of exactly two \velo tracks. 
The efficiency is found to be the same down to the $10^{-3}$ level. This stability is expected since the \spd requirement is very loose and therefore no systematic uncertainty is attributed to this source.

The relative difference between calculating the photon-related trigger efficiency with only $\eta$ or \piz mesons is set as a systematic uncertainty.
This uncertainty accounts for possible biases arising from the fit procedure, including those related to the signal and background modelling.

The \JPsiee sample is normalised using the cross-section measured for $\jpsi\to\mup\mun$ production in UPC~\cite{LHCb-PAPER-2022-012},  
therefore the total uncertainty of the latter measurement is considered as a systematic uncertainty here.\footnote{The luminosity uncertainty is excluded as the same data are used in this search.} 
The resulting relative systematic uncertainty on the \JPsiee cross-section is 4\%.

As the \starlight event generator is used to model the \ALP candidates, while the \superchic generator is used to compute the cross-sections entering the determination of the lower limits on the \ALP coupling, a potential bias may arise from differences between the two generators. 
To quantify this effect, the \ALP signal acceptance within the LHCb detector is compared using both generators. 
A difference of 31\% is observed, with \superchic exhibiting a larger acceptance than \starlight. 
This difference is symmetrised and assigned as a systematic uncertainty on the signal normalisation.

No systematic uncertainty is assigned to the cross-sections calculated using \superchic that are used to normalise both the signal and background, since previous measurements with larger data samples have shown good agreement with these predictions~\cite{CMS:2024bnt,ATLAS:2020hii}.
However, for the determination of the ggF cross-section, the conservative treatment of soft-survival effects in PbPb collisions~\cite{Kaidalov:2003vg,Goncalves:2010dw} leads to different predictions compared to Ref.~\cite{Levin:2008gi} and Ref.~\cite{Basso:2017mue}, where both apply a Glauber model~\cite{Miller:2007ri} to PbPb interactions.
These two predictions are used to define an upper and lower bound for the normalisation factor. 

\begin{table}[bt]
\caption{\small Relative systematic uncertainties considered for the limit calculation.
Where a  value spans multiple columns, it indicates the uncertainty is correlated between samples.}\label{tab:sys}
\renewcommand{\arraystretch}{1.25}
\begin{center}\begin{tabular}{l|c|c|c|c|c}
\hline
Process & \ALP & \LbL & \ggF & \BWh & \JPsiee \\ \hline \hline
Luminosity & \multicolumn{5}{c}{$\pm4.2\%$} \\
\hline
\herschel efficiency  & \multicolumn{5}{c}{$\pm1.4\%$} \\
\hline
Photon efficiency & \multicolumn{5}{c}{$\pm4.5\%$}  \\ 
\hline
Multiplicity requirements  & \multicolumn{3}{c|}{$\pm0.3\%$} & $^{\,+\,39\%}_{\,-\,28\%}$ & $\pm50\%$ \\
\hline
Photon trigger efficiency  & \multicolumn{5}{c}{$^{\,+\,5.4\%}_{\,-\,6.7\%}$} \\ 
\hline
\JPsiee cross-section & \multicolumn{4}{c|}{$-$}  & $\pm 4\%$ \\
\hline
\ALP simulation  & $\pm31\%$ & \multicolumn{4}{c}{$-$} \\
\hline
Soft survival & \multicolumn{2}{c|}{$-$} & $^{\,+\,46\%}_{\,-\,11\%}$ & \multicolumn{2}{c}{$-$} \\ 
\hline
\hline
Total & $\pm32\%$ & $^{\,+\,8.3\%}_{\,-\,9.2\%}$ & $^{\,+\,47\%}_{\,-\,14\%}$ & $^{\,+\,40\%}_{\,-\,30\%}$ & $\pm51\%$ \\
\hline
\end{tabular}
\end{center}
\end{table}

\section{Results and conclusions}
\label{sec:results} 

\begin{figure}[tb]
\begin{center}
\includegraphics[width=0.95\linewidth]{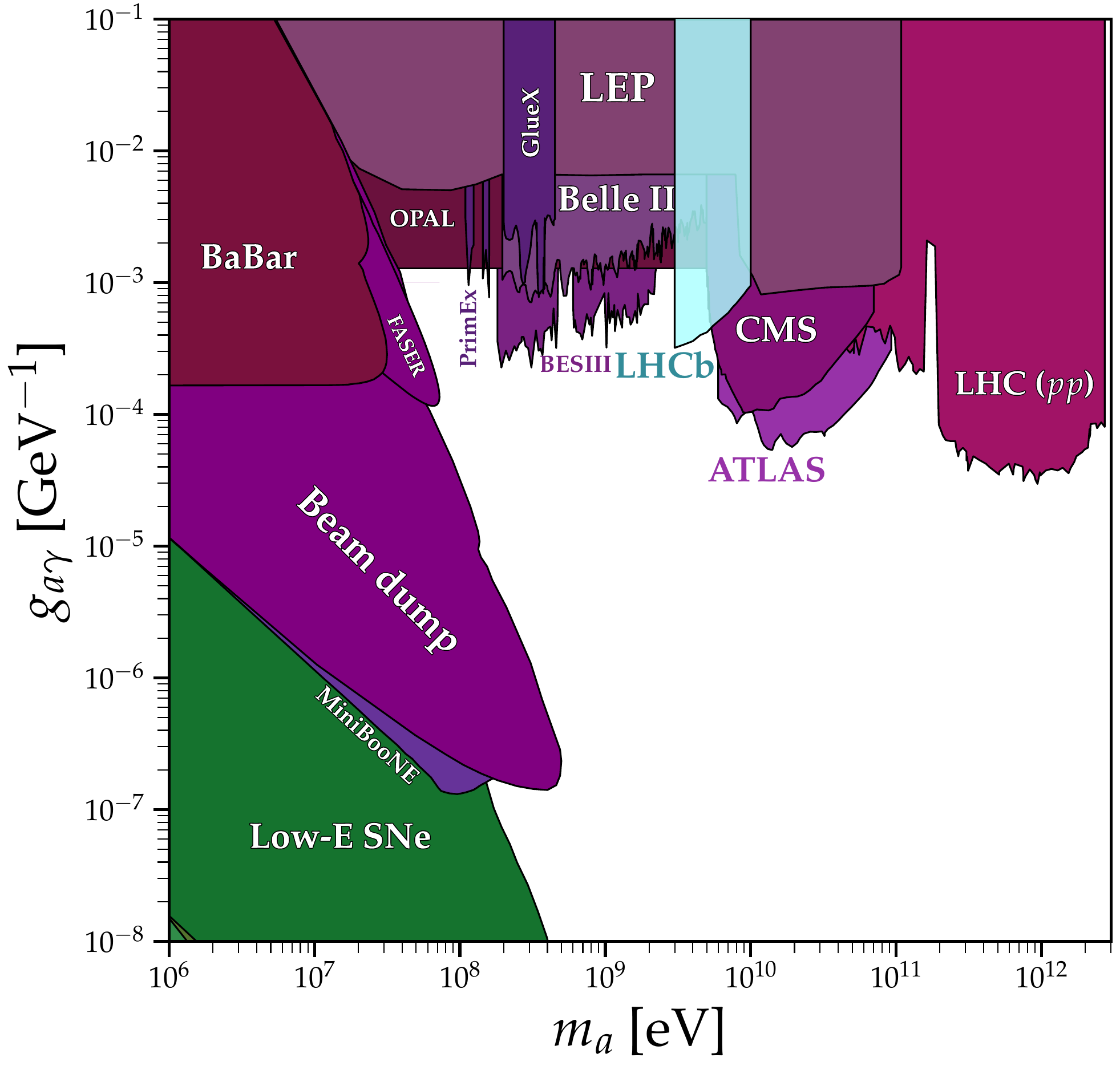}
\vspace*{-0.5cm}
\end{center}
\caption{Collider, accelerator and beam dump bounds with  \lhcb exclusion (cyan) at 95\%~CL in the axion-photon coupling \galp versus axion mass \malp plane, assuming \ALPs only couple to photons. 
Previous limits have been obtained from ATLAS~\cite{ATLAS:2020hii},  CMS~\cite{CMS:2018erd,CMS:2024bnt}, 
LEP~\cite{OPAL:2002vhf,Knapen:2016moh,Jaeckel:2015jla}, \belletwo~\cite{Belle-II:2020jti}, beam dumps \cite{beamdump, beamdump2}, \besiii \cite{BESIII}, PrimEx \cite{PrimEx:2010fvg,Aloni:2019ruo}, GlueX \cite{GlueX}, LHC $pp$ \cite{lhcpp1, lhcpp2, lhcpp3, ATLAS:2023zfc}, FASER \cite{FASER}, BaBar \cite{Babar} and low energy supernovae constraints \cite{Low-E}.}
\label{fig:limcomp}
\end{figure}

\begin{figure}[htb]
\begin{center}
\includegraphics[width=0.70\linewidth]{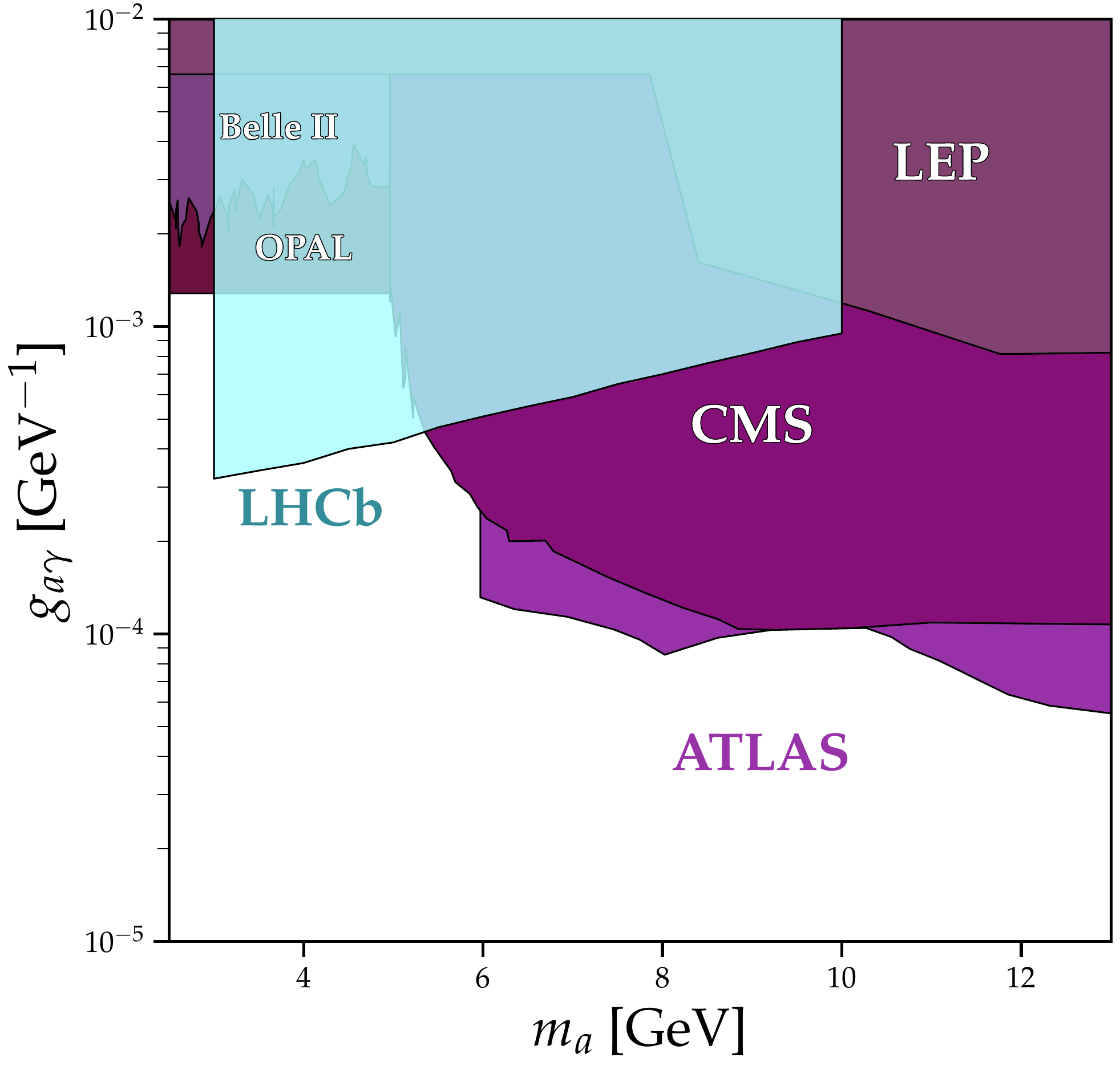}
\vspace*{-0.5cm}
\end{center}
\caption{Zoomed view of the LHCb exclusion limits at $95\%$ CL in the axion-photon coupling \galp versus axion mass \malp plane, assuming \ALPs only couple to photons. 
The figure also shows exclusion limits obtained from ATLAS~\cite{Aad:2020cje}, CMS~\cite{CMS:2018erd,CMS:2024bnt}, LEP~\cite{Knapen:2016moh} and Belle II~\cite{Belle-II:2020jti}.}
\label{fig:limcomp_zoom}
\end{figure}

As no diphoton candidates are observed in this analysis, limits are set on the \ALP model  implemented in \superchic.
The limits are determined with the $CL_{s}$ method as implemented in the pyhf repository~\cite{Cowan:2010js,pyhf_joss}.
A single overall systematic uncertainty is used to account for all sources described in Sec.~\ref{sec:systematic}. Statistical uncertainties from the size of simulated samples used are included in the limit calculation.
The resulting upper limits with 95\% $CL_{s}$ on the \ALP production cross-section vary between 1.1\pb and 3.1\pb for masses between 3 and 10\gev.
The corresponding limits on the coupling \galp vary between $3.2\times10^{-4}\gev^{-1}$ and  $9.5\times10^{-4}\gev^{-1}$ over the same mass interval.
The exclusion region set by this paper is compared to other experiments in Figs.~\ref{fig:limcomp} and \ref{fig:limcomp_zoom} using the AxionLimits package~\cite{AxionLimits}. 

In summary, this analysis uses data corresponding to an integrated luminosity of 0.228\invnb collected in 2018 at a nucleon-nucleon centre-of-mass energy of $\sqsnn = 5.02\tev$.
It represents the first search based on the production of two photons in the pseudorapidity interval $2.0 < \eta(\gamma) < 4.0$ in PbPb collisions.  
These data allow  the most stringent limits to be set to date on \ALP production from photon-photon fusion in the mass range 3--5\gev, complementing existing bounds from $e^+e^-$ colliders.

\section*{Acknowledgements}
%
%
\noindent We express our gratitude to our colleagues in the CERN
accelerator departments for the excellent performance of the LHC. We
thank the technical and administrative staff at the LHCb
institutes.
We acknowledge support from CERN and from the national agencies:
ARC (Australia);
CAPES, CNPq, FAPERJ and FINEP (Brazil); 
MOST and NSFC (China); 
CNRS/IN2P3 and CEA (France);  
BMFTR, DFG and MPG (Germany);
NKFIH (Hungary);              
INFN (Italy); 
NWO (Netherlands); 
MNiSW and NCN (Poland); 
MEC/IFA (Romania); 
MICIU and AEI (Spain);
SNSF and SER (Switzerland); 
NASU (Ukraine); 
STFC (United Kingdom); 
DOE NP and NSF (USA).
We acknowledge the computing resources that are provided by ARDC (Australia), 
CBPF (Brazil),
CERN, 
IHEP and LZU (China),
IN2P3 (France), 
KIT and DESY (Germany), 
INFN (Italy), 
SURF (Netherlands),
Polish WLCG (Poland),
IFIN-HH (Romania), 
PIC (Spain), CSCS (Switzerland), 
GridPP (United Kingdom),
and NSF (USA).  
We are indebted to the communities behind the multiple open-source
software packages on which we depend.
Individual groups or members have received support from
RTP (Australia), 
FWO Odysseus grant G0ASD25N (Belgium), 
Key Research Program of Frontier Sciences of CAS, CAS PIFI, CAS CCEPP (China); 
Minciencias (Colombia);
EPLANET, Marie Sk\l{}odowska-Curie Actions, ERC and NextGenerationEU (European Union);
A*MIDEX, ANR, IPhU and Labex P2IO, and R\'{e}gion Auvergne-Rh\^{o}ne-Alpes (France);
Alexander-von-Humboldt Foundation (Germany);
ICSC (Italy); 
Severo Ochoa and Mar\'ia de Maeztu Units of Excellence, GVA, XuntaGal, GENCAT, InTalent-Inditex and Prog.~Atracci\'on Talento CM (Spain);
the Leverhulme Trust, the Royal Society and UKRI (United Kingdom).

\addcontentsline{toc}{section}{References}

\bibliographystyle{LHCb/LHCb}
\bibliography{main,LHCb/standard,LHCb/LHCb-PAPER,LHCb/LHCb-CONF,LHCb/LHCb-DP,LHCb/LHCb-TDR}

\newpage
\centerline
{\large\bf LHCb collaboration}
\begin
{flushleft}
\small
R.~Aaij$^{39}$\lhcborcid{0000-0003-0533-1952},
M.~Abdelfatah$^{71}$,
A.S.W.~Abdelmotteleb$^{59}$\lhcborcid{0000-0001-7905-0542},
C.~Abellan~Beteta$^{53}$\lhcborcid{0009-0009-0869-6798},
F.~Abudin\'en$^{61}$\lhcborcid{0000-0002-6737-3528},
T.~Ackernley$^{63}$\lhcborcid{0000-0002-5951-3498},
A.A.~Adefisoye$^{71}$\lhcborcid{0000-0003-2448-1550},
B.~Adeva$^{49}$\lhcborcid{0000-0001-9756-3712},
M.~Adinolfi$^{57}$\lhcborcid{0000-0002-1326-1264},
P.~Adlarson$^{87,44}$\lhcborcid{0000-0001-6280-3851}
% ...

{\footnotesize \it

% ...
$^{39}$Nikhef National Institute for Subatomic Physics, Amsterdam, Netherlands\\
% ...
$^{44}$National Center for Nuclear Research (NCBJ), Warsaw, Poland\\
% ...
$^{49}$Instituto Galego de F{\'\i}sica de Altas Enerx{\'\i}as (IGFAE), Universidade de Santiago de Compostela, Santiago de Compostela, Spain\\
% ...
$^{53}$Physik-Institut, Universit{\"a}t Z{\"u}rich, Z{\"u}rich, Switzerland\\
% ...
$^{57}$H.H. Wills Physics Laboratory, University of Bristol, Bristol, United Kingdom\\
% ...
$^{59}$Department of Physics, University of Warwick, Coventry, United Kingdom\\
% ...
$^{61}$School of Physics and Astronomy, University of Edinburgh, Edinburgh, United Kingdom\\
$^{62}$School of Physics and Astronomy, University of Glasgow, Glasgow, United Kingdom\\
$^{63}$Oliver Lodge Laboratory, University of Liverpool, Liverpool, United Kingdom\\
% ...
$^{71}$Syracuse University, Syracuse, NY, United States\\
% ...
$^{87}$Department of Physics and Astronomy, Uppsala University, Uppsala, Sweden, associated to $^{62}$\\
% ...
}
\end{flushleft}

\end{document}